\documentclass[%
 reprint,
 superscriptaddress,
 amsmath,amssymb,
 aps,
floatfix,
]{revtex4-2}

\usepackage{graphicx}
\usepackage{dcolumn}
\usepackage{bm}
\usepackage{hyperref}
\usepackage{gensymb}
\usepackage{xcolor}
\usepackage{algpseudocode}
\usepackage{algorithm}
\usepackage{lipsum}  

\hypersetup{
    colorlinks=true,
    linkcolor=blue,
    citecolor=blue, 
    urlcolor=black
    }

\begin{document}

\title{Proton Radiography Inversions with Source Extraction and Comparison to Mesh Methods}


\author{J. Griff-McMahon}
\email{jgriffmc@pppl.gov}
\affiliation{Department of Astrophysical Sciences, Princeton University, Princeton, New Jersey 08544, USA}
\affiliation{Princeton Plasma Physics Laboratory, Princeton, New Jersey 08540, USA}

\author{V. Valenzuela-Villaseca}
\affiliation{Department of Astrophysical Sciences, Princeton University, Princeton, New Jersey 08544, USA}

\author{S. Malko}
\affiliation{Princeton Plasma Physics Laboratory, Princeton, New Jersey 08540, USA}

\author{G. Fiksel}
\affiliation{Center for Ultrafast Optical Science, University of Michigan, Ann Arbor, Michigan 48109, USA}

\author{M. J. Rosenberg}
\affiliation{Laboratory for Laser Energetics, University of Rochester, Rochester, New York 14623, USA}

\author{D. B. Schaeffer}
\affiliation{Department of Physics and Astronomy, University of California Los Angeles, Los Angeles, California 90095, USA}

\author{W. Fox}
\affiliation{Department of Astrophysical Sciences, Princeton University, Princeton, New Jersey 08544, USA}
\affiliation{Princeton Plasma Physics Laboratory, Princeton, New Jersey 08540, USA}

\begin{abstract}
Proton radiography is a central diagnostic technique for measuring electromagnetic (EM) fields in high-energy-density, laser-produced plasmas. 
In this technique, protons traverse the plasma where they accumulate small EM deflections which lead to variations in the proton fluence pattern on a detector. Path-integrated EM fields can then be extracted from the fluence image through an inversion process.
In this work, experiments of laser-driven foils were conducted on the OMEGA laser and magnetic field reconstructions were performed using both ``fluence-based"  techniques and high-fidelity ``mesh-based" methods. We implement nonzero boundary conditions into the inversion and show their importance by comparing against mesh measurements. Good agreement between the methods is found only when nonzero boundary conditions are used. We also introduce an approach to determine the unperturbed proton source profile, which is a required input in fluence reconstruction algorithms. In this approach, a fluence inversion is embedded inside of a mesh region, which provides overconstrained magnetic boundary conditions. A source profile is then iteratively optimized to satisfy the boundary information. This method substantially enhances the accuracy in recovering EM fields. Lastly, we propose a scheme to quantify uncertainty in the final inversion that is introduced through errors in the source retrieval.
\end{abstract}

\maketitle

\section{Introduction}
Proton radiography is the prevailing diagnostic used to measure electromagnetic (EM) fields in high-energy-density (HED) laser-produced plasma experiments \cite{mackinnon_proton_2004,li_measuring_2006,fox_filamentation_2013,rosenberg_slowing_2015,schaeffer_generation_2017,schaeffer_proton_2023}. In this method, protons with MeV energies are sent through the plasma and weakly deflected by the EM fields. The protons can either be generated as a beam by target normal sheath acceleration (TNSA) \cite{zylstra_using_2012} or as an isotropic source generated by an implosion of a D$^3$He-filled capsule. The proton fluence is then measured on a detector, and the intensity pattern is used to infer information about the path-integrated EM fields in the plasma with high spatial (10 to 50 $\mu$m) and temporal resolution (10 to 150 ps)  \cite{manuel_source_2012}. In addition to valuable qualitative information, several methodologies have been developed over the past decade to extract quantitative 2D images of the path-integrated EM fields from the measured proton fluence images \cite{kugland_invited_2012,bott_proton_2017,graziani_inferring_2017,campbell_magnetic_2020,davies_quantitative_2023,fox_proton_2023}. These inversions have directly supported our understanding of magnetic reconnection \cite{fox_fast_2020,pearcy_experimental_2024}, magnetized collisionless shocks \cite{schaeffer_direct_2019,li_collisionless_2019}, and magnetic field generation in laser-produced plasmas \cite{campbell_magnetic_2020,heuer_diagnosing_2022}.

One of the main drawbacks in proton inversion algorithms is that they require knowledge of the source fluence pattern (i.e. what the proton fluence image would look like in the absence of fields). Presently, there is no way to measure the source fluence pattern and the deflected fluence pattern simultaneously. In practice, the source pattern is often estimated by smoothing the measured proton image with a low-pass filter \cite{pearcy_experimental_2024,bott_proton_2017}. However, this approach is sub-optimal as modulations in the observed proton fluence can stem from both EM field structures and variations in the source pattern. In particular, protons emitted from a D$^3$He backlighter capsule can exhibit variations in the proton fluence $\gtrsim$ 50\% in a single shot on the detector, although the variation is typically small over solid angles $\leq$ 1.1 deg \cite{manuel_source_2012}. Low-pass filtering to estimate the source is most vulnerable to missing fields that vary over large spatial scales which could easily be mistaken as variations in the source itself.
Recent progress has shown that statistical properties of the source can be leveraged to retrieve an estimate of the path-integrated fields without knowledge of the exact source profile \cite{kasim_retrieving_2019}. This approach generates an ensemble of path-integrated fields, each from a different potential source profile, which enables statistical analyses. Nevertheless, the exact source profile and inversion remains unknown in this technique.

In addition, the results from inversion algorithms are strongly dependent on boundary conditions as local changes at the boundaries can influence the entire inversion. Boundary conditions are especially important for topics such as magnetic reconnection, where the absolute value of the field is of interest (not just quantities such as RMS or local field variations). Boundary values can even be used to directly constrain the source pattern in 1-dimensional inversions \cite{fox_proton_2023,fox_fast_2020}. Despite their impact on the inversion, nonzero boundary conditions are often not considered. For example, \textcite{palmer_field_2019} and \textcite{tubman_observations_2021} used vanishing boundary conditions in their inversions, despite mesh distortion at the edge of the domain that implies the presence of EM fields. Furthermore, widely used reconstruction codes (e.g. the PROBLEM code \cite{bott_proton_2017} in the version as of this writing) apply zero-deflection boundary conditions that prevent protons from being deflected into or out of the domain. This is equivalent to enforcing that the corresponding electric and magnetic field components, $E_\perp$ and $B_\parallel$, vanish along the boundary. Vanishing boundary conditions may lead to distorted EM reconstructions if the boundary is close to the interaction or if there are extended EM fields as measured for example in \cite{griff-mcmahon_measurements_2023}. Although inversion techniques have been validated in isolation (with e.g. forward model comparisons), the ambiguity in the source profile and assumptions in the boundary conditions introduce uncertainty in the accuracy of the inversion under real experimental conditions. 

There is another class of EM field measurement, often referred to as ``proton deflectometry", that directly measures proton deflections using a mesh, rather than inverting the proton fluence pattern. In this scheme, a mesh splits the incident proton beam into a grid of beamlets, whose deflection can be tracked by observing changes to the final position of individual beamlets on the detector \cite{li_measuring_2006,petrasso_lorentz_2009}. The D$^3$He backlighter capsule also emits x-rays which can be measured simultaneously to provide a reference for the unperturbed proton trajectories \cite{malko_design_2022,johnson_proton_2022}. The path-integrated field is then calculated in a single shot from a shift between the undeflected trajectory (x-rays) and the deflected trajectory (protons). This scheme has been benchmarked against known field profiles \cite{johnson_proton_2022} and the x-ray fiducials have demonstrated their importance for high-fidelity reconstructions and for capturing extended field measurements \cite{griff-mcmahon_measurements_2023}. In general, ``mesh-based" reconstructions of EM fields offer better accuracy but worse spatial resolution than fluence-based inversions.

In this work, experiments on magnetic field generation from laser-driven foils were conducted and used to benchmark proton radiography inversions against high-fidelity, mesh measurements. Specialized laser targets were fielded, designed to allow both techniques in a single experimental shot. These ``hybrid" shots have both fluence and mesh deflection information, which can be combined to improve the inversion accuracy. Building on Ref. \cite{fox_proton_2023}, which considered the application of boundary condition data in 1-D reconstructions, here we extend the technique to 2-D reconstructions. In particular, the mesh measurements provide boundary conditions for the inversion and can be leveraged to determine the source profile, addressing a long-standing problem in proton radiography inversions. In this scheme, the source is iteratively optimized to simultaneously match both tangential and normal components of the proton boundary deflection informed from the mesh. We find good agreement between the inversion and mesh measurements only when nonzero magnetic boundary conditions are used and the source is optimized. We also introduce a method to estimate the uncertainty in inversions introduced through errors in the source retrieval.

The text is organized as follows. Section \ref{sec:MA_scheme} provides an overview of the Monge-Amp\`ere inversion scheme used throughout the paper. We then introduce nonzero boundary conditions into the algorithm and validate the implementation using forward modeling. Section \ref{sec:Exp} discusses the experiment and offers a comparison of the magnetic reconstructions using both the mesh and fluence methods. In section \ref{sec:InvSource}, we present a scheme to extract the source profile and apply it to an experimental shot. Lastly, section \ref{sec:InvErr} develops a method to estimate the error in an inversion that originates from uncertainty in the source identification.

\section{EM Field Reconstructions in the Monge-Amp\`ere scheme} \label{sec:MA_scheme}
\subsection{Overview of the Monge-Amp\`ere Scheme}
There are many techniques to reconstruct path-integrated EM fields directly from proton fluence images \cite{kugland_invited_2012,bott_proton_2017,graziani_inferring_2017,campbell_magnetic_2020,davies_quantitative_2023,fox_proton_2023}. The analysis in this work focuses on the Monge-Amp\`ere (M-A) relaxation algorithm \cite{sulman_efficient_2011}, as implemented in the PROBLEM code \cite{bott_proton_2017}. This is a nonlinear, finite-difference solver that is second-order in space. However, the various techniques we introduce and discuss can be applied to other inversion algorithms more generally. This section outlines the framework of the M-A algorithm, introduces nonzero boundary conditions, and validates the implementation.

\begin{figure} [b]
	\includegraphics[width=\linewidth]{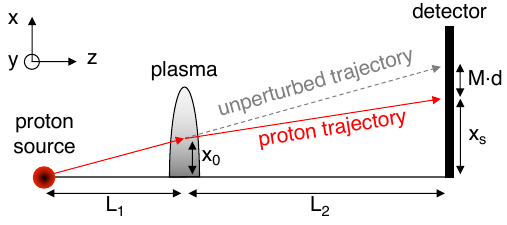}
	\caption{Schematic of proton radiography.}
	\label{fig:PradSetup}
\end{figure}

The general set-up of proton radiography is shown in Fig. \ref{fig:PradSetup}. A source provides MeV-level protons which traverse the plasma and accumulate small-angle deflections from the EM fields. After exiting the probed EM field region, the protons propagate ballistically to a detector. The analysis of proton images is often aided by several approximations, including the paraxial, point-projection, and small-deflection approximations. These are discussed in detail in the work of  \textcite{bott_proton_2017}. In our experiments, the paraxial parameter $\delta \alpha=0.3$, the point-projection parameter $\delta \beta =0.01$, and the maximum proton deflection angle is $0.01$ rad, justifying the use of these approximations.

Under these approximations, a proton passing through initial coordinate $\bm{x}_0$ in the plasma plane is mapped to the detector coordinate $\bm{x}_s$ according to Eq. \eqref{Eqn:ProtonDeflection}.
\begin{equation}
    \bm{x}_s(\bm{x}_0) = M\bm{x}_0 + \frac{L_2}{v_p} \bm{w}(\bm{x}_0)
    \label{Eqn:ProtonDeflection}
\end{equation}
Here, $M=(L_1+L_2)/L_1$ is the image magnification, $L_1$ is the distance from the source to the plasma, $L_2$ is the distance from the plasma to the detector, $v_p$ is the proton speed, and $\bm{w}(\bm{x}_0)$ is the perpendicular deflection velocity \cite{kugland_invited_2012} given by Eq. \eqref{eq:PerpDefVel}.
\begin{equation}
   \bm{w}(\bm{x}_0)=\frac{e}{m_p}\int ds \left[\hat{z}\times\bm{B}(\bm{x}(s)) + \frac{\bm{E_\perp}(\bm{x}(s))}{v_p}\right].
   \label{eq:PerpDefVel}
\end{equation}
The deflection velocity depends on the proton charge-to-mass ratio and the transverse electric and magnetic fields that the proton is subjected to along its trajectory through the plasma $\bm{x}(s)$. Often, it is helpful to work in the plasma plane coordinates, which demagnifies the final particle position to $\bm{x}_f$.
\begin{align}
    \bm{x}_f(\bm{x}_0) &= \bm{x}_0 + \bm{d}(\bm{x}_0)\\
    \bm{d}(\bm{x}_0)&=\frac{L_2}{M v_p} \bm{w}(\bm{x}_0) \label{eq:defVec}
\end{align}
The goal of fluence inversions is then reduced to determining a 2D map of proton deflections $\bm{d}(\bm{x}_0)$. 

The measured fluence on the detector screen $\Psi(\bm{x}_0)$ relies on the principle of local fluence conservation \cite{kugland_invited_2012}. It is determined from the source fluence $\Psi_0(\bm{x}_0)$ and the Jacobian of the  mapping from the unperturbed coordinate $\bm{x}_0$ to the final coordinate $\bm{x}_f$. 
\begin{equation}
    \Psi(\bm{x}_f)=\frac{\Psi_0(\bm{x}_0)}{\det\left[\nabla_{\perp0}(\bm{x}_f(\bm{x}_0)\right]}
    \label{eq:kugland}
\end{equation} 

Due to the irrotational nature of the proton deflection, the final coordinate $\bm{x}_f$ can be recast as the gradient of a potential $\phi$, which formulates Eq. \eqref{eq:kugland} into a Monge-Amp\`ere equation \cite{kugland_invited_2012}.
\begin{align}
    \bm{x}_f(\bm{x}_0) &= \nabla_{\perp0} \phi(\bm{x}_0)\\
    \Psi(\nabla_{\perp0}\phi(\bm{x}_0))&=\frac{\Psi_0(\bm{x}_0)}{\det\left[\nabla_{\perp0} \nabla_{\perp0} \phi(\bm{x}_0)\right]} \label{eq:MAE}
\end{align}

The Monge-Amp\`ere equation has been studied extensively and plays a significant role in transport theory and geometrical optics \cite{gangbo_geometry_1996}. Eq. \eqref{eq:MAE} is solved efficiently by finding the steady-state solution of the logarithmic-parabolic Monge- Amp\`ere equation \cite{sulman_efficient_2011} shown in Eq. \eqref{Eqn:LPMAE}. Here, a time variable $t$ is introduced to evolve the potential $\phi$ toward its steady-state solution. Note that this variable has no physical bearing and is purely used as a tool in the inversion algorithm.
\begin{equation}
    \frac{\partial \phi}{\partial t} = \log \left[ \frac{\Psi[\nabla _{\perp0} \phi(\bm{x}_0)] \det [\nabla_{\perp0} \nabla_{\perp0} \phi(\bm{x}_0)]}{\Psi_0(\bm{x}_0)} \right]
    \label{Eqn:LPMAE}
\end{equation}
This is a nonlinear reconstruction algorithm that accurately reconstructs fluence images in the nonlinear injective regime. In this regime, proton deflections measured on the detector have a similar size to the field coherence length after magnification \cite{bott_proton_2017,kugland_invited_2012}.

The path-integrated EM fields are then converted from the potential $\phi$ as shown in Eq. \eqref{eq:FieldFromDeflection}. The electric field contribution is often dropped since $v_pB\gg E$ in many systems. Accordingly, the electric contribution is ignored for most of this paper.
\begin{equation}
    \int d\bm{l} \times \bm{B} = \frac{m_p M v_p}{e L_2} \left[ \nabla_{\perp 0} \phi(\bm{x}_0)-\bm{x}_0 \right]
    \label{eq:FieldFromDeflection}
\end{equation}
At its core, however, the inversion solves for proton deflections which can be attributed to electric fields or magnetic fields or some combination of the two.

\subsection{Nonzero Boundary Conditions}
Inversion algorithms typically adopt vanishing EM field boundary conditions which enforce that proton deflections normal to the boundary (corresponding to tangential magnetic fields) are zero. This conserves total proton count inside the inversion domain. However, if the field region is not well separated from the edge, using vanishing boundary conditions is incorrect and compromises the accuracy of the reconstruction. In the PROBEM code \cite{bott_proton_2017} (as of the time of this writing), vanishing normal boundary deflections are implemented with Neumann boundary conditions.
\begin{equation}
    \nabla_\perp \phi\cdot \hat{\bm{n}} = \bm{x}_0 \cdot \hat{\bm{n}}
\end{equation}

To the best of our knowledge, nonzero boundary conditions have never been considered for the M-A scheme, although they have been considered for 1D codes (see e.g. \cite{fox_proton_2023}) as well as 2D linear solvers \cite{graziani_inferring_2017}. Here, we introduce a proton deflection normal to the boundary $d_\perp(\bm{x}_0,t)$ into the M-A scheme so that the boundary condition becomes.

\begin{equation}
    \nabla_\perp \phi\cdot \hat{\bm{n}} = \bm{x}_0 \cdot \hat{\bm{n}} + d_\perp(\bm{x}_0,t).
    \label{Eqn:NonVansishBC}
\end{equation}

We find that the edge deflection must be slowly ``turned on" over many time steps to keep the system in approximate equilibrium. This maintains monotonicity of $\bm{x}_f$ and prevents proton crossings known as caustics which jeopardize the reconstruction and prevent a unique determination of the path-integrated fields. If the boundary condition is turned on too quickly, then edge deflections may overrun the neighboring deflections and produce irreversible caustics in the solve. The slow turn-on is implemented by linearly increasing $d_\perp(\bm{x}_0,t)$ to its final value $d_{\perp0}(\bm{x}_0)$ over a time period $T$, whereupon the code continues to run until steady-state convergence criteria is met.
\begin{equation}
    d_\perp(\bm{x}_0,t) = d_{\perp0}(\bm{x}_0) \times \begin{cases}
    t/T & t<T\\
    1 & t\geq T
\end{cases}
\label{eq:normDisp}
\end{equation}
We note that only perpendicular deflection $d_\perp$ can be constrained while parallel deflection $d_\parallel$ must remain free so as not to overconstrain the PDE system.

A nuanced aspect of including nonzero boundary conditions is that global proton number may not be conserved within the original domain. With nonzero boundary conditions, the domain of the undeflected image $\Omega_0$ is mapped to a new domain in the deflected image: $\Omega_f = \nabla_\perp \phi(\Omega_0)$. As a result, proton count is typically not conserved in the original domain $\Omega_0$. For most inversion algorithms, this would require a careful treatment of the source proton fluence (see e.g. \cite{fox_proton_2023}). However, Eq. \eqref{Eqn:LPMAE} is insensitive to proton conservation as well as uniform scalings of the source profile. To demonstrate, consider a source profile $\Psi_0(\bm{x}_0)$ that conserves proton count and scale it uniformly by a factor $A$ in Eq. \eqref{Eqn:LPMAE}.
\begin{equation}
    \frac{\partial \phi}{\partial t} = \log \left[ \frac{K}{A\Psi_0(\bm{x}_0)} \right] = \log \left[ \frac{K}{\Psi_0(\bm{x}_0)} \right] - \log{A} 
        \label{eq:PotentialDrift} 
\end{equation}
Here, $K$ is the numerator in Eq. \eqref{Eqn:LPMAE}. 
\begin{equation}
    K=\Psi[\nabla _{\perp0} \phi(\bm{x}_0)] \det [\nabla_{\perp0} \nabla_{\perp0} \phi(\bm{x}_0)]
\end{equation}

The scaling factor introduces a constant term that uniformly changes $\phi$ as the solution is progressed. Fortuitously, this term has no spatial dependence and does not contribute to the spatial derivatives of $\phi$. This result is a consequence of the M-A relaxation technique. In other algorithms, different absolute fluence profiles lead to significantly different inversions, for example, as shown in the 1-D reconstruction algorithm in Ref. \cite{fox_proton_2023}. Due to this characteristic of the M-A scheme, scaled versions of the same source will result in identical inversions. The first term on the RHS has a mean value which vanishes \cite{sulman_efficient_2011}, but the second term does not and may lead to numerical rounding errors if $\phi$ gets very large. However, subtracting off the mean value of $\phi$ every few steps prevents this from happening.

Finally, there is a subtle difference between inward and outward edge deflections that arises from the interpolation term $\Psi(\bm{x}_f)$ in Eq. \eqref{Eqn:LPMAE}. This term samples the proton image inside the deformed domain $\Omega_f$. If the deflection is inward, $\bm{x}_f$ lies within the original domain $\Omega_0$ so the interpolation is standard. However, outward deflection samples $\bm{x}_f$ from outside of the original domain and can pose issues. Small outward deflections may be treated similarly to inward deflections and extrapolated outside of the domain, either by nearest value or linear extrapolation. Larger deflections require a more robust solution in which two regions are examined: the first region is used for the inversion as usual while the second is larger and is used for the interpolation step. Accounting for this outward deflection is particularly relevant for geometries with defocusing polarities. The geometries discussed in this work have focusing polarities so the boundary deflection is always inwards.

\subsection{Validation of Nonzero Boundary Conditions}
\begin{figure}
	\includegraphics[width=\linewidth]{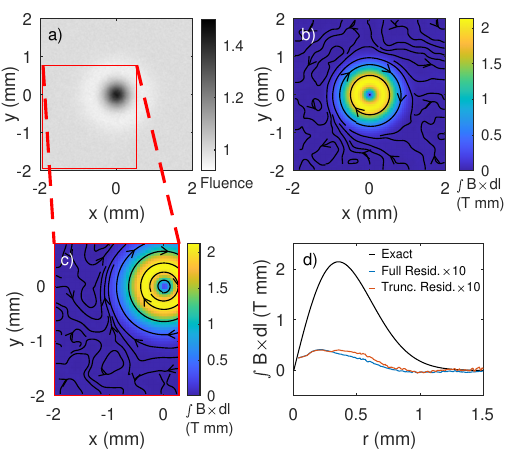}
	\caption{\textbf{Synthetic Radiograph and Magnetic Reconstruction.} a) Synthetic radiograph in object coordinates. b) Magnetic reconstruction using the full domain and vanishing magnetic boundary conditions. c) Magnetic reconstruction using a truncated domain with cutoff at $x=0.25$ mm and $y=0.75$ mm and nonzero boundary conditions from the analytic form. d) Radial profile of the path-integrated magnetic field for the analytic form and error residuals scaled by $10\times$.}
	\label{fig:FwdModelValidation}
\end{figure}

The implementation of nonzero boundary conditions was validated using a forward model technique. A synthetic radiograph is generated by propagating $4\times10^{8}$ simulated protons through a focusing annular profile given by Eq. \eqref{Eqn:Bdl_analytic} with scale length $a=0.5$ mm. This profile is qualitatively representative of the magnetic fields that are generated around the laser spot when a laser irradiates a solid target \cite{li_measuring_2006}. The synthetic detector has a spatial resolution of 10 $\mu$m in the target plane. 

\begin{equation}
    \int B_\phi(r)\ dz = 5\ \frac{r}{a} \exp \left[{-\left( \frac{r}{a} \right)^2} \right] \text{ T}\ \text{mm}
    \label{Eqn:Bdl_analytic}
\end{equation}

The synthetic radiograph is shown in Fig. \ref{fig:FwdModelValidation}a and was inverted two times with different domains to validate the implementation. The first reconstruction, shown in Fig. \ref{fig:FwdModelValidation}b, used the full domain of the radiograph ($-2$ to 2 mm in both dimensions) such that the magnetic field decayed to less than $10^{-6}$ of its maximum value at the boundaries. In this reconstruction, vanishing boundary conditions were used. The second reconstruction is shown in Fig. \ref{fig:FwdModelValidation}c and used a truncated domain depicted by the black rectangle in Fig. \ref{fig:FwdModelValidation}a. The $x$ domain spans from $-2$ to $0.25$ mm and the $y$ domain spans from $-2$ to $0.75$ mm, resulting in nonzero magnetic fields along the top and right boundaries. In this reconstruction, nonzero boundary conditions were implemented from the tangential component of the analytic magnetic field.

Both reconstructions successfully converged to within a root-mean-square error (RMSE) of 0.02 T$\,$mm and maximum error of 0.07 T$\,$mm. These correspond to 1\% and 3.5\% of the peak field strength, respectively. The averaged radial profile, shown in Fig. \ref{fig:FwdModelValidation}d, confirms that the errors in both reconstructions are almost identical and at the 1\% level. These may both be improved with better spatial resolution to resolve the numerical derivatives more smoothly and with more protons in the forward model to reduce noise in the synthetic radiograph. Overall, these results demonstrate the robustness of the technique to nonzero boundary conditions in the inversion.

Conversely, if the truncated region incorrectly uses \emph{vanishing} boundary conditions (not shown), the reconstruction has a large RMSE of 0.44 T$\,$mm and maximum error of 2 T$\,$mm. That is to say, the maximum error is at the 100\% level since the boundary cuts through the maximum in the field profile. The magnitude of the errors highlight the importance of using correct boundary conditions on the global solution. 

\section{Application to Experiment} \label{sec:Exp}
\subsection{Experimental Setup}
\begin{figure} [b]
	\includegraphics[width=\linewidth]{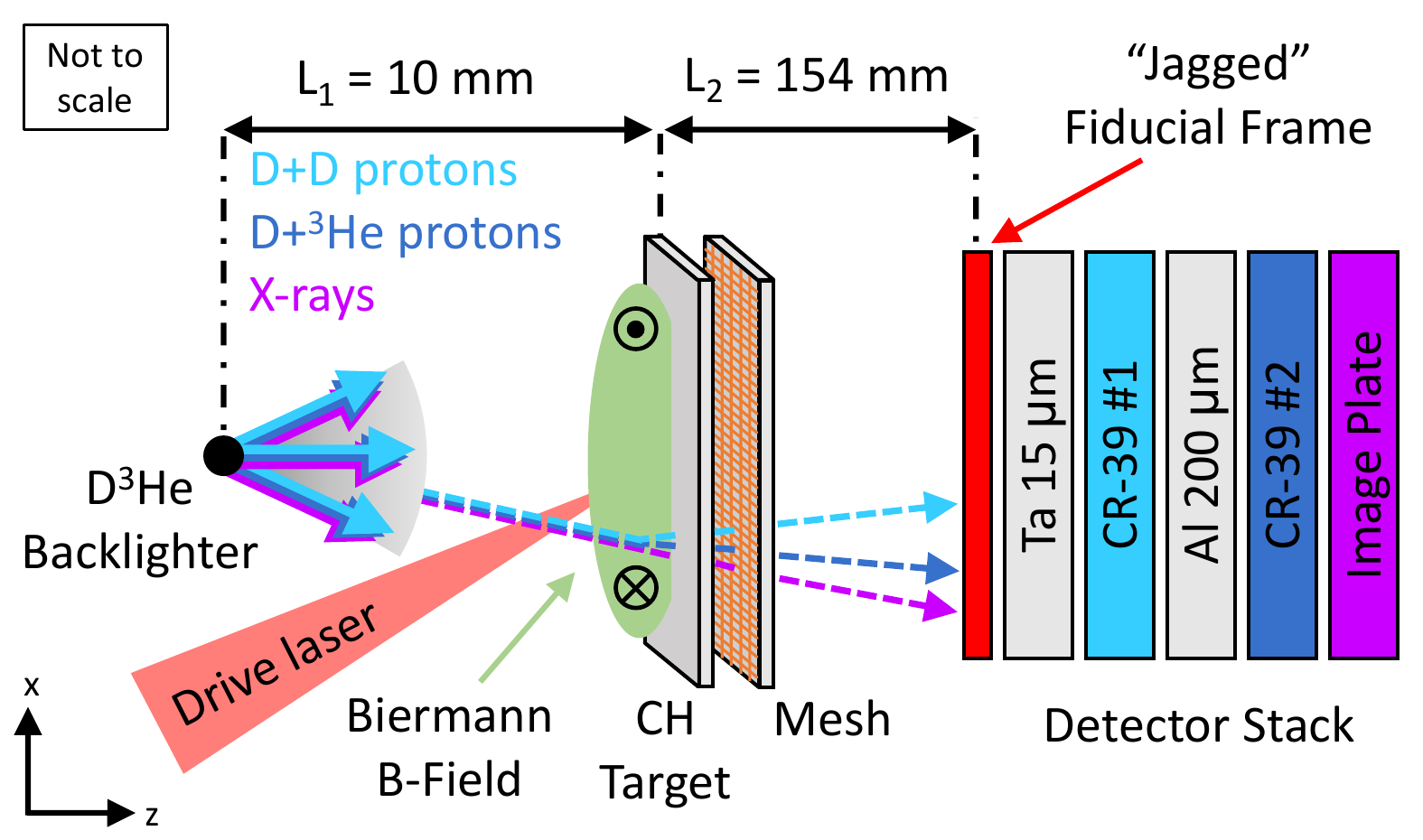}
	\caption{\textbf{Experimental set up.} Biermann-battery magnetic fields are imaged by protons and x-rays. The D$+^3$He protons have a birth-energy of 14.7 MeV and imaged the fields at $t=+1.4$ ns.}
	\label{fig:ExperimentalSetup}
\end{figure}

\begin{figure*}
    \includegraphics[width=\textwidth]{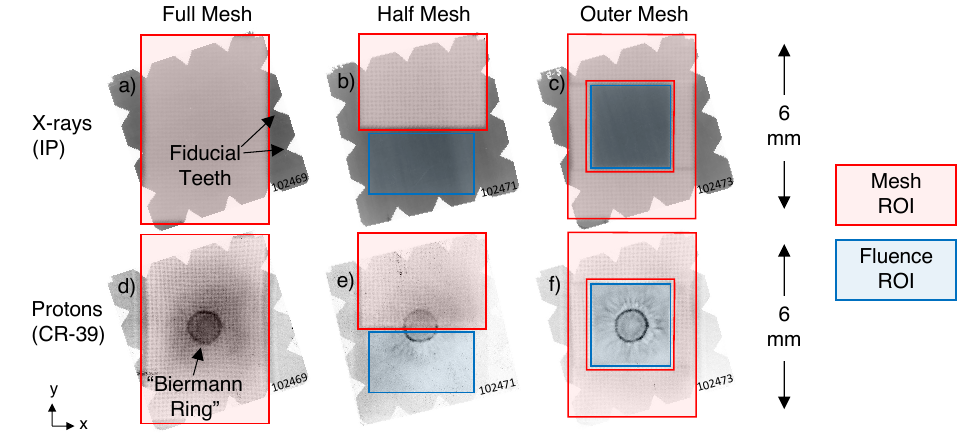}
    \caption{\textbf{Aligned and rotated x-ray and proton images}. The top row (a-c) shows aligned images x-ray fluence and the bottom row (d-f) show images of proton fluence (darker regions received higher fluence). A different mesh configuration is shown in each column. The red and blue areas indicate selected regions for mesh reconstruction (red) and fluence inversions (blue). The images are rotated so that beamlet rows are horizontal.}
    \label{fig:Combined_IP_CR-39}
\end{figure*}

A series of experiments were performed using the Omega Laser Facility at the Laboratory for Laser Energetics at the University of Rochester. As shown in Fig. \ref{fig:ExperimentalSetup}, two overlapped laser beams impinged on a 25 $\mu$m-thick, 6.6$\times$5 mm CH foil to produce a single expanding plume and generate magnetic fields from the Biermann-battery effect \cite{biermann_uber_1950,li_measuring_2006}. The lasers delivered 1 kJ of total energy within a 1 ns pulse duration. The spatial profile was a supergaussian with exponent 5.2 and  $1/e$ waist of 358 $\mu$m from an SG5 Distributed Phase Plate \cite{lin_distributed_1995}, resulting in a peak laser intensity of $3\times10^{14}$ W/cm$^2$. 

Additionally, 25 other beams were used to drive a 420~$\mu$m diameter capsule filled with D$^3$He gas to produce a source of protons and x-rays. These imaged the magnetic fields generated on the surface of the foil. Two monoenergetic populations of protons are emitted from both the D+D and D$+^3$He fusion reactions. However, the analysis presented here primarily focuses on the D$+^3$He protons, which have a fusion birth energy of 14.7 MeV and probed the fields at $t=+$1.4 ns after the drive lasers turned on.

A nickel mesh was attached to the rear surface of the foil to break the incident protons and x-rays into a grid of beamlets. The mesh was 60 $\mu$m thick and had a spatial period of 150 $\mu$m. The beamlets are used to directly measure proton deflection. Three different mesh configurations were fielded in the experiment to test mesh and fluence reconstructions from the same shot. The mesh configurations are shown in Fig. \ref{fig:Combined_IP_CR-39} and discussed in more detail in the next section (\ref{subsec:results}).

After passing through the target, the protons and x-rays travel to a detector stack where the proton fluence pattern was detected on CR-39 and the x-ray pattern was measured on an image plate. A jagged fiducial frame at the front of the detector stack imprinted shadows of fiducial ``teeth" along the edges of the x-ray and proton images which were used to align the images for the mesh deflection measurements \cite{johnson_proton_2022,malko_design_2022}.

\subsection{Experimental Results} \label{subsec:results}
Experimentally measured x-ray and proton images are shown in Fig. \ref{fig:Combined_IP_CR-39}. These images used nominally the same laser parameters to generate comparable EM field features. The images have been rotated $\sim15$ deg so that the rows of the mesh grid are horizontal. Each column in Fig. \ref{fig:Combined_IP_CR-39} shows a different mesh configuration with the x-ray image along the top row and proton image along the bottom row for the same shot. Each pair of x-ray and proton images was aligned to each other using the fiducial teeth. An edge detection algorithm was used to minimize the alignment uncertainty to 0.4 pixels, corresponding to $0.5$ T$\,$mm of path-integrated magnetic field strength at 16.4$\times$ magnification. After the images are aligned, pixels in the proton image map directly to pixels in the x-ray image. Displacement from x-ray beamlet to proton beamlet extracts a measurement of the proton deflection induced by EM fields \cite{johnson_proton_2022,malko_design_2022}.
\begin{equation}
    \label{eqn:beamlet_def}
    \int d\mathbf{l} \times \mathbf{B} = \frac{m_p v_p}{e L_2} (\bm{x}_s - M\bm{x}_0)
\end{equation}
Here, $\bm{x}_s - M\bm{x}_0$ is the beamlet deflection vector in the detector plane. In these shots, the D+D neutron yield varied from 1.3--1.8$\times 10^{9}$ and the average proton yield was $\sim$100 counts per pixel.

The left column (Fig. \ref{fig:Combined_IP_CR-39}a,d) used a mesh that covered the entire foil resulting in an entirely mesh-based reconstruction within the red rectangular region. The center column (Fig. \ref{fig:Combined_IP_CR-39}b,e) used a mesh that covered only the top half of the foil so the bottom half could be reconstructed using fluence techniques (blue rectangle). This configuration enables a comparison of mesh reconstructions and fluence inversions from the same shot. And the last column (Fig. \ref{fig:Combined_IP_CR-39}c,f) used a mesh with a square cutout in the center, that supports an inversion that is embedded inside a mesh reconstruction. This unique mesh design provides high-fidelity, magnetic boundary conditions to the inversion. The ``Biermann Ring" in the proton images is a common feature in radiographs of laser-driven foils and approximately corresponds to the location of peak magnetic field strength around the laser spot.

\begin{figure*}
    \includegraphics[width=\linewidth]{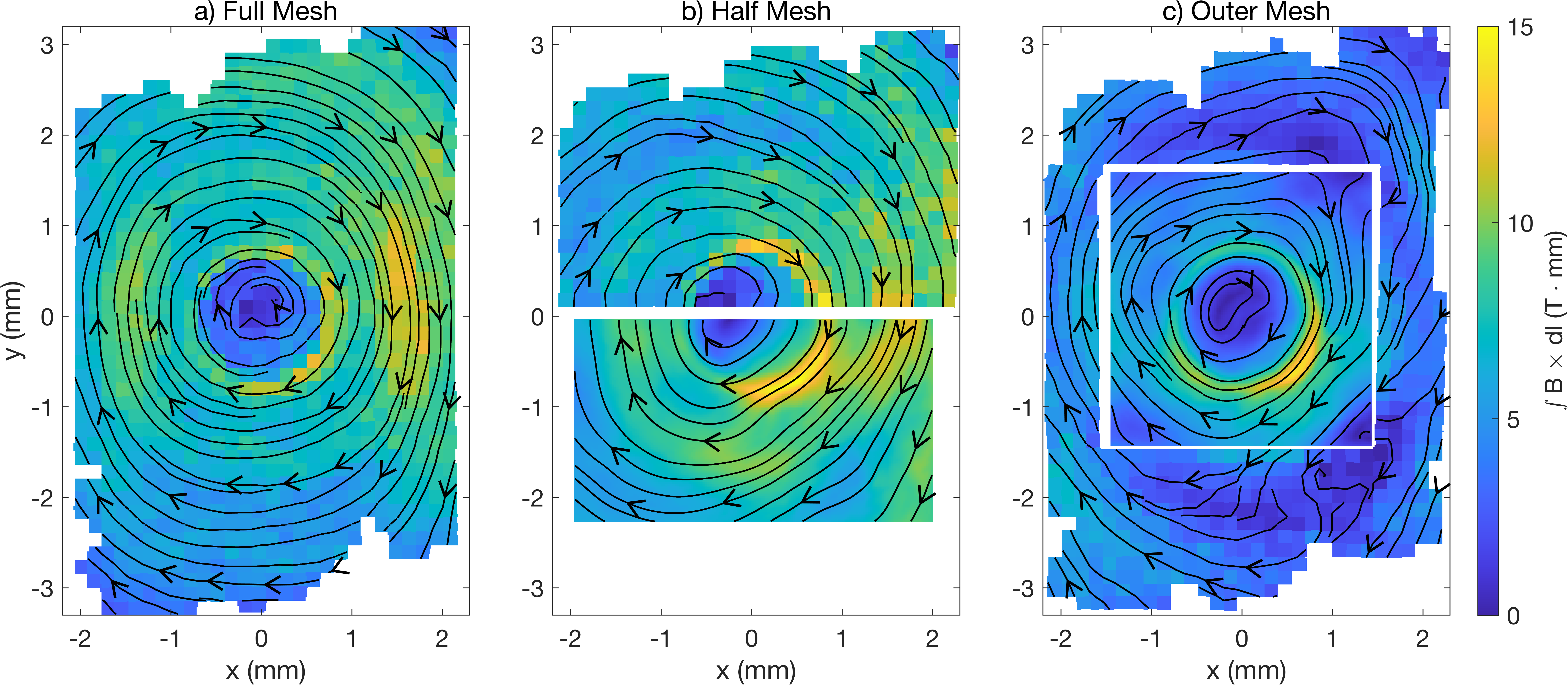}
    \caption{\textbf{Magnetic field reconstruction}: Magnetic field extracted from mesh deflection and fluence inversions in different mesh configurations. a) The full mesh case was reconstructed entirely using mesh techniques. b) The top region of the half mesh used mesh while the bottom region used fluence methods. c) The outer region of the outer mesh configuration used mesh, whereas the inner region used fluence methods. The tangential component of the magnetic boundary conditions in the inversion are informed from the mesh measurements. The electric fields are assumed negligible.}
    \label{fig:ReconstructedBField}
\end{figure*}

Figure \ref{fig:ReconstructedBField} shows magnetic reconstructions of the different targets using mesh and fluence techniques. In each configuration, a clockwise toroidal magnetic field is measured around the laser spot with a null on-axis. The half mesh and outer mesh inversions used non-zero magnetic boundary conditions informed from the mesh region (i.e. comparison of the beamlet deflections between the x-ray and proton images). In the half mesh configuration, symmetry over the x-axis is assumed so that the tangential component of the magnetic field along the boundary, $B_\parallel$, is given by the nearest mesh reconstruction point in the corresponding top half of the image. Indeed, the field structure is expected to be azimuthally symmetric under ideal conditions. In this case, the boundary conditions have high uncertainty as they are not directly measured along three sides. In the outer mesh configuration, $B_\parallel$ along the boundary was simply taken to be the nearest mesh point. The reconstructions assume that proton deflections are entirely due to magnetic fields. Electric contributions are discussed in Appendix \ref{App:EvB}.

The mesh measurements provide information about the perpendicular and parallel components of proton deflection along the boundary of the inversion. These can be cast as Neumann and Dirichlet boundary conditions, respectively. However, after considering the structure of the PDE in Eq. \eqref{Eqn:LPMAE} and assuming a specified source profile, it is evident that \textit{only} one type of boundary condition (Neumann or Dirichlet) can be applied to the PDE solution. Applying both would lead to an over-determined system that does not have a solution in general. The resolution is that the equivalence (and overdetermination) between Dirichlet and Neumann conditions assumes that the PDE is being solved with the exact and correct proton source profile, $\Psi_0$.

Indeed, we find that enforcing the ``primary” Neumann condition (perpendicular deflection) with an un-optimized source profile produces a solution that generally does not match the \textit{directly measured} parallel deflections. This indicates a fundamental connection between boundary conditions and the observed and source proton profiles. To address this, we develop a framework to incorporate this information on equal footing. Namely, the key detail to reconcile the mismatch of this ``secondary" boundary condition is our ignorance of the source profile which adds a degree of freedom. By tailoring the source profile, the secondary boundary condition can also be satisfied. Moreover, tailoring the source to match both boundary conditions extracts the true source profile in the absence of an experimentally measured one. The details of the source extraction are described in Section \ref{sec:InvSource}.

Figures \ref{fig:ReconstructedBField}(b) and \ref{fig:ReconstructedBField}(c) show fluence reconstructions for the half-mesh and outer mesh configurations using this analysis technique. In both configurations, we determine a source profile which satisfies both magnetic boundary conditions to within 1 T $\,$mm, as evidenced by the continuity of the field lines across the mesh-fluence border. 

\begin{figure}
    \includegraphics[width=\linewidth]{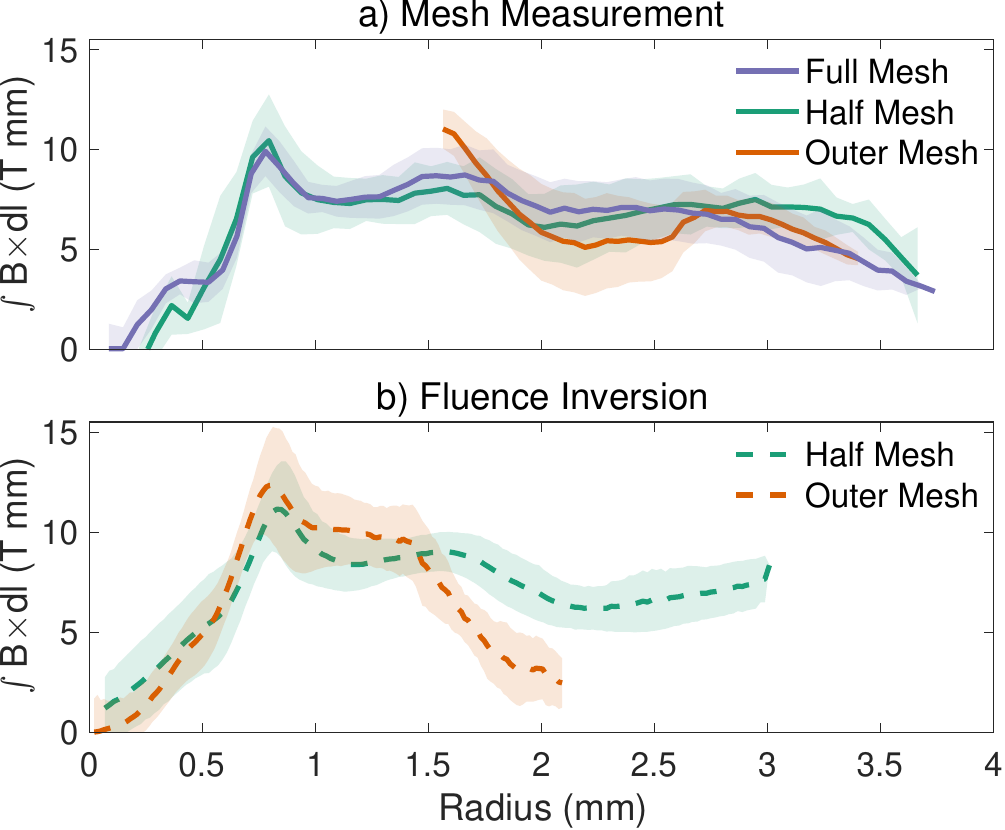}
    \caption{\textbf{Magnetic field profile}: Radial profile of the path-integrated toroidal magnetic field, averaged in the azimuthal direction for a) mesh measurements and b) fluence inversions. The various target configurations are shown in different colors and the shaded region shows the standard deviation within each radial bin. In the outer mesh configuration, the deflection due to electric fields has been subtracted out as described in Appendix \ref{App:EvB}.}
    \label{fig:Bth_radial}
\end{figure}

To provide a quantitative analysis between mesh and fluence reconstructions, the azimuthal component of the field is extracted and radially binned to produce a radial lineout shown in Fig. \ref{fig:Bth_radial}. Different targets are displayed in different colors and the inversion technique is separated between Figs. \ref{fig:Bth_radial}a and \ref{fig:Bth_radial}b. The shaded regions illustrate the standard deviation in each radial bin. The outer mesh configuration has been corrected to account for transverse electric fields using the method proposed in \cite{griff-mcmahon_measurements_2023}. Further details of the analysis can be found in Appendix \ref{App:EvB}. The electric fields in the other configurations were relatively weak and did not require this correction. 

For radii less than 1.5 mm, the different techniques and targets show excellent agreement and have the same averaged peak field strength to within 25\%. In addition, the full mesh reconstruction (shown in blue) and both reconstructions of the half mesh (shown in green) align well, even capturing the multiple peak structures at $r=$ 0.8 and 1.6 mm. 

However, there is a notable discrepancy in the field profiles in the outer mesh case, shown in orange in Fig. \ref{fig:Bth_radial}. The fluence inversion dips below the other configurations between $r=$ 1.5 and 2.2 mm. These radii correspond to regions of the radiograph which contain both fluence and mesh data, particularly the corners of the fluence region. The field strength also differs between the two techniques in this region.  This discrepancy is likely a consequence of imperfect electric field subtraction as discussed in Appendix \ref{App:EvB}. To summarize the findings of this Appendix, we infer there was a line-integrated electric field up to $400~$MV$\,$m$\,$mm$^{-1}$ in the outward direction. This field leads to an equivalent deflection of 2~mm in the detector plane, which is comparable in magnitude to the total deflection of 3~mm of deflection, but in the opposite direction; the B fields are oriented to deflect protons inward while the electric fields are oriented to deflect protons outward.

\section{Source Profile Extraction} \label{sec:InvSource}
\subsection{Method of Source Extraction}
Uncertainty in the undeflected source profile has undermined proton radiography inversions since their inception. A correct choice of source profile is important for faithful reconstructions, but is difficult to obtain as the source and deflected proton images cannot be measured simultaneously. In this section, we demonstrate the importance of obtaining the correct source profile and introduce a method to solve for the correct profile by simultaneously satisfying both magnetic field boundary conditions.

As shown in Fig. \ref{fig:ReconstructedBField}, it is possible to embed a fluence inversion inside of a mesh reconstruction. In this setup, edge values of $B_\parallel$ and $B_\perp$ are extracted from the mesh region. We recall the inversion will automatically match the specified $B_\parallel$ along the boundary as that is enforced in the Neumann conditions of the inversion algorithm. However, we find that depending on the source profile (which we are solving for), $B_\perp$ may or may not match at the boundary. We assert that we can optimize for the correct fluence profile that also matches $B_\perp$ along the boundary.

To begin, we represent the source $\Psi_0$ as a sum of Chebyshev basis polynomials.
\begin{equation} \label{eq:chebyBasis}
    \Psi_0(x,y) = \sum_{i,j=0}^{N-1} A_{ij}T_i(x)T_j(y)
\end{equation}
Here, $A_{ij}$ is an element in the weight array $A$, $T_i$ is the $i^{th}$ order Chebyshev polynomial of the first kind, and $x$ and $y$ are normalized coordinates spanning from $-1$ to 1. It is convenient to enforce $A_{00}=1$, corresponding to the uniform fluence coefficient. In these shots, we find the other elements are on the order of 0.01 to 0.1. Before solving, the source is re-normalized to the fluence level in the proton image to reduce the drift in the potential shown in Eq. \eqref{eq:PotentialDrift}. We find that using 4 Chebyshev polynomials ($0^{th}$ to $3^{th}$ orders) is typically sufficient to reduce the boundary residual of $B_\perp$ to acceptable levels of $\sim 1$ T$\ $mm. The low order of the weight array allows only large-scale fluctuations in the source (as measured in \cite{manuel_source_2012}) and speeds up convergence.

In the first step of the optimization, the weight array $A$ is initialized to a uniform fluence so that $A_{ij}=\delta_{i0}\delta_{j0}$. Then, successive inversions calculate the boundary error of $B_\perp$ as each element in $A$ is independently adjusted by the finite difference step size $\lambda$ as shown in Eq. \eqref{eq:errorAdjust}. The error, $\epsilon$, is quantified as the root-mean-square-error (RMSE) between the inversion and closest mesh point along the boundary. The derivative of $\epsilon$ for the $j^{th}$ element of $A$ is estimated through a centered difference approximation according to Eq. \eqref{eq:FiniteDiff}. 
\begin{align}
    \epsilon_j^{\pm}&=\epsilon(A;A_j\pm \lambda) \label{eq:errorAdjust}\\
    \frac{\partial \epsilon}{\partial j} &\approx \frac{\epsilon_j^+ - \epsilon_j^+}{2 \lambda} \label{eq:FiniteDiff}
\end{align}

\begin{figure*}
    \includegraphics[width=\linewidth]{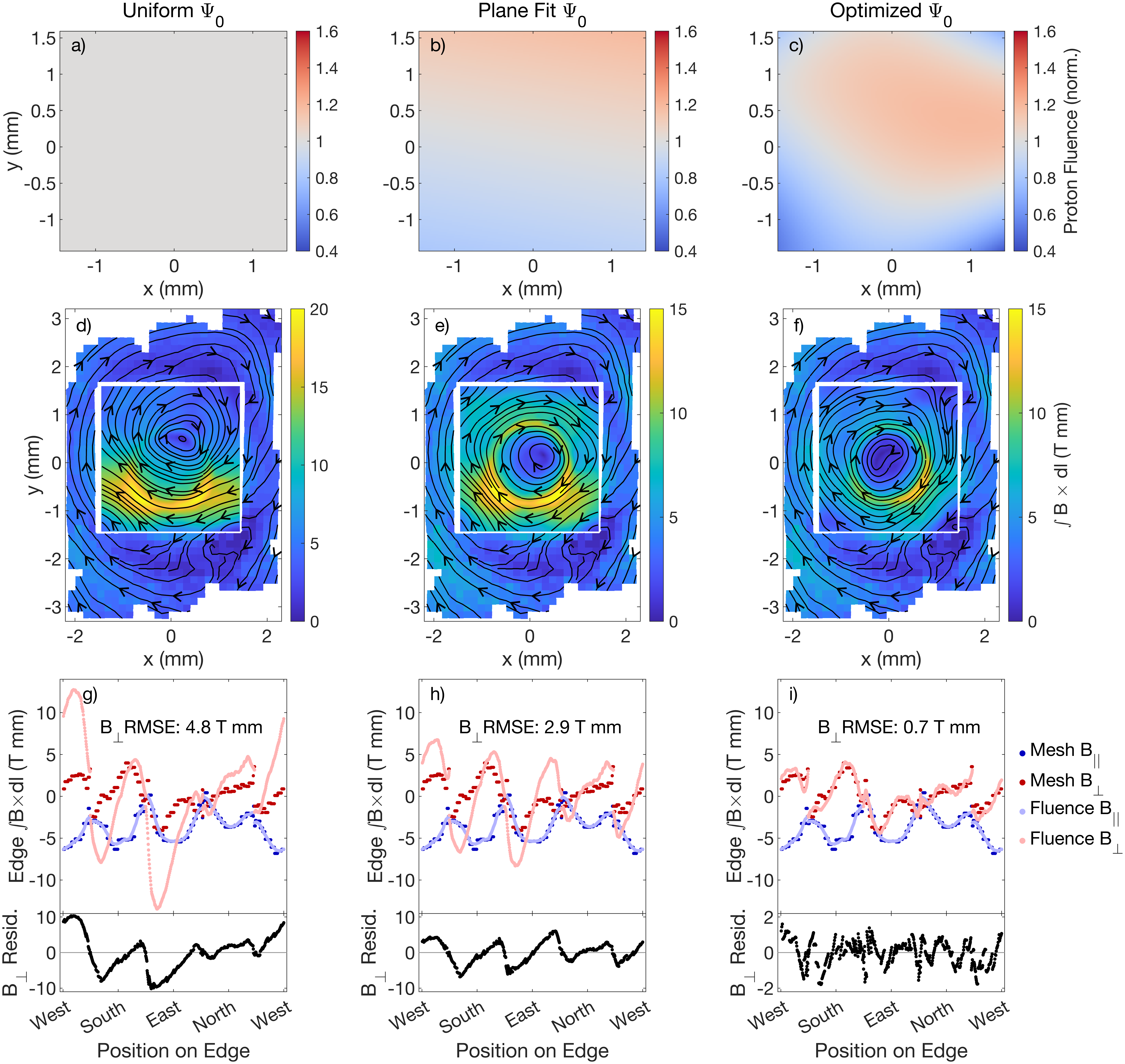}
    \caption{\textbf{Source profile effects}:
    \emph{Top row}: Spatial profile of source $\Psi_0$ used in the reconstruction, normalized to the mean value. a) Uniform $\Psi_0$. b) Plane fit to the proton image. c) Optimized $\Psi_0$ to match the mesh $B_\perp$ at the boundary.
    \emph{Middle row}: Embedded inversion using different background proton fluence profiles $\Psi_0$.
    \emph{Bottom row}: Boundary values of the tangential and normal magnetic fields for mesh and fluence reconstructions. $B_\parallel$ boundary conditions in the inversion are smoothed. The subplot below shows the residual in $B_\perp$ between inversion and mesh reconstruction. All deflection is assumed to be of magnetic origin.
    }
    \label{fig:SourceOpt}
\end{figure*}

After calculating the error derivative for each element of $A$, a new weight array is estimated using a simple gradient descent method. 
\begin{equation}
    A_{n+1}=A_n - \gamma \nabla_j \epsilon
    \label{eq:GradDesc}
\end{equation}
Here, $\gamma$ is the gradient descent step size.

This process is iterated until the error falls below a determined error threshold or the error converges. It is possible that the chosen basis does not permit sources that fall below the desired error. In this case, a higher order Chebyshev basis or different basis set may be chosen. It is helpful to scale down the proton image to an array of size $\sim50\times50$ to speed up the source optimization. This does not significantly affect the optimized source, as small scale-structure is already excluded due to the smallness of the basis set. Once the source is converged, the image and source can both be scaled to full-size. An outline of the optimization algorithm is found in Appendix \ref{App:GradDescAlg}.

\subsection{Source Extraction of Experimental Data}
Figure \ref{fig:SourceOpt} shows the results of magnetic field reconstruction with increasingly optimized source fluence. The outer mesh configuration is inverted three times, each with a different source profile to highlight the influence of the sources on the resulting inversion. The different sources are shown along the top row of Fig. \ref{fig:SourceOpt}, the magnetic inversions are embedded in the middle row, and a boundary value comparison is presented in the bottom row.

The left column uses a uniform source (Fig. \ref{fig:SourceOpt}a) to generate the inversion (Fig. \ref{fig:SourceOpt}d). Visual inspection reveals that the field is strongly asymmetric and the field lines are extremely mismatched across the inversion-mesh interface. Figure \ref{fig:SourceOpt}g shows both components of the magnetic boundary values for the mesh measurement (dark blue and red) and the inner inversion (light blue and red). In the inversion, $B_\parallel$ was constrained to fit the nearest mesh points after smoothing, which explains the agreement between the light and dark blue lines. However, $B_\perp$ was free and serves as a metric for the accuracy of the source profile when compared to the directly measured edge values. In this case, the values of $B_\perp$ strongly deviate from the mesh values, resulting in a large RMSE of 4.8 T$\,$mm, which is a significant error on the order of the peak field strength and over twice the RMS field over the measured region. In particular, several locations in the inversion obtain a field strength 2 to 3 times higher than what the mesh method measures and are therefore in significant error. The residual of edge values of $B_\perp$ are shown in the bottom half of Fig. \ref{fig:SourceOpt}g and indicate systematic regions of edge mismatch. The high error and discontinuity of the field lines suggest that a uniform $\Psi_0$ is a poor choice of source profile.

The central column uses a source profile that is a plane fit to the observed proton fluence pattern (Fig. \ref{fig:SourceOpt}b). This method is often used in experimental inversions and offers insight into the inversion errors introduced in analyses without boundary information. Relative to the uniform $\Psi_0$ case, the corresponding reconstruction in Fig. \ref{fig:SourceOpt}e has better symmetry and field-line matching across the boundary, although there are still clear regions along the edge that have large discontinuities in $B_\perp$. Figure \ref{fig:SourceOpt}h shows the $B_\perp$ RMSE is reduced to 2.9 T$\,$mm and the local regions of high peak field have been eliminated.

Finally, the right column uses an optimized source profile that minimizes the $B_\perp$ RMSE. A 4$\times$4 weight array was used to generate the source using Chebyshev basis polynomials from Eq. \eqref{eq:chebyBasis}. This led to smooth variations across the domain as shown in Fig. \ref{fig:SourceOpt}c. The optimized source has qualitative similarities to the plane fit $\Psi_0$; the top of the domain has greater fluence than the bottom. The main differences lie in the corners of the optimized $\Psi_0$, which have less fluence than the plane fit. The optimized $\Psi_0$ also has large variations in the fluence, ranging from normalized values of 0.4 to 1.2. Variations of this magnitude are also found in a null shot of the D$^3$He backlighter, lending credibility to the large source variation. The null shot is described in Appendix \ref{App:nullShot}. The field lines in the inversion almost perfectly match across the boundary (Fig. \ref{fig:SourceOpt}f). This is supported in Fig. \ref{fig:SourceOpt}i which shows that the edge fields in mesh and fluence methods are almost identical and have a $B_\perp$ RMSE of 0.7 T$\,$mm. Additionally, the $B_\perp$ residuals are largely uncorrelated and indicate systematic mismatching has been minimized.

The optimized source profile took 13 iterations for the source to converge and used finite difference step size $\lambda=0.02$ and gradient descent step size $\gamma=0.02$. Each iteration required 30 inversions (15 array elements $\times$ 2 solves per element) and the entire optimization took 30 minutes on a personal laptop, demonstrating the feasibility of this approach. The fast run time was aided by starting the Monge-Amp\`ere solve from a previous solution $\phi(\bm{x}_0)$, to reduce the number of required iterations, and by downscaling the proton image by 50\%.

As shown above, the basis set of Eq. \eqref{eq:chebyBasis} is successful at significantly improving the match of the edge magnetic fields between the fluence reconstruction and direct mesh measurements, reducing the residual errors from the 100\%-level to erros that are comparable to individual beamlet measurement errors. Therefore, this method makes significant steps over prior techniques.

Nevertheless we should re-iterate the fundamental assumptions of this method and possible blind spots. Most fundamentally, our method implicity assumes that the source fluence profile variation is entirely at long wavelengths that can be captured by the basis set of Eq. \eqref{eq:chebyBasis}. By extension, we assumes that all high-frequency components in the observed fluence profile are produced by true EM fields in the plasma. This assumption is supported by direct measurements of the fluence variations (Ref. \cite{manuel_source_2012} and repeated in Appendix B for the present data).

\section{Inversion Error} \label{sec:InvErr}

Understanding and evaluating the errors in proton radiography inversions is crucial to obtaining quantitiative physics results. Two main sources of uncertainty that can enter fluence-based inversions are: (1) error in the boundary information which propagates into the interior of the inversion and (2) error in selecting the correct source profile. \textcite{kasim_retrieving_2019} has demonstrated progress in the latter through a statistical procedure by which a large ensemble of potential source profiles are generated via a Gaussian process. The ensemble relies on several experimental null shots to understand the characteristic correlation lengths and magnitudes of the source perturbations. The sources are then used to create a family of inversions where the spread in the inversion fields estimates the uncertainty introduced from ignorance in the exact source profile. However, this work did not incorporate boundary information; $B_\parallel$ vanishes on the edge of the domain and the influence of the source on boundary values of $B_\perp$ is not considered. Since boundary information of the fields is known from beamlet deflections in the mesh, then the possible source profiles are limited to only those that satisfy these boundaries, to within some error tolerance.  In this section, we outline a Monte Carlo scheme to quantify the inversion error that stems from uncertainty in the source profile when boundary information is known.

\begin{figure} [h!]
    \includegraphics[width=\linewidth]{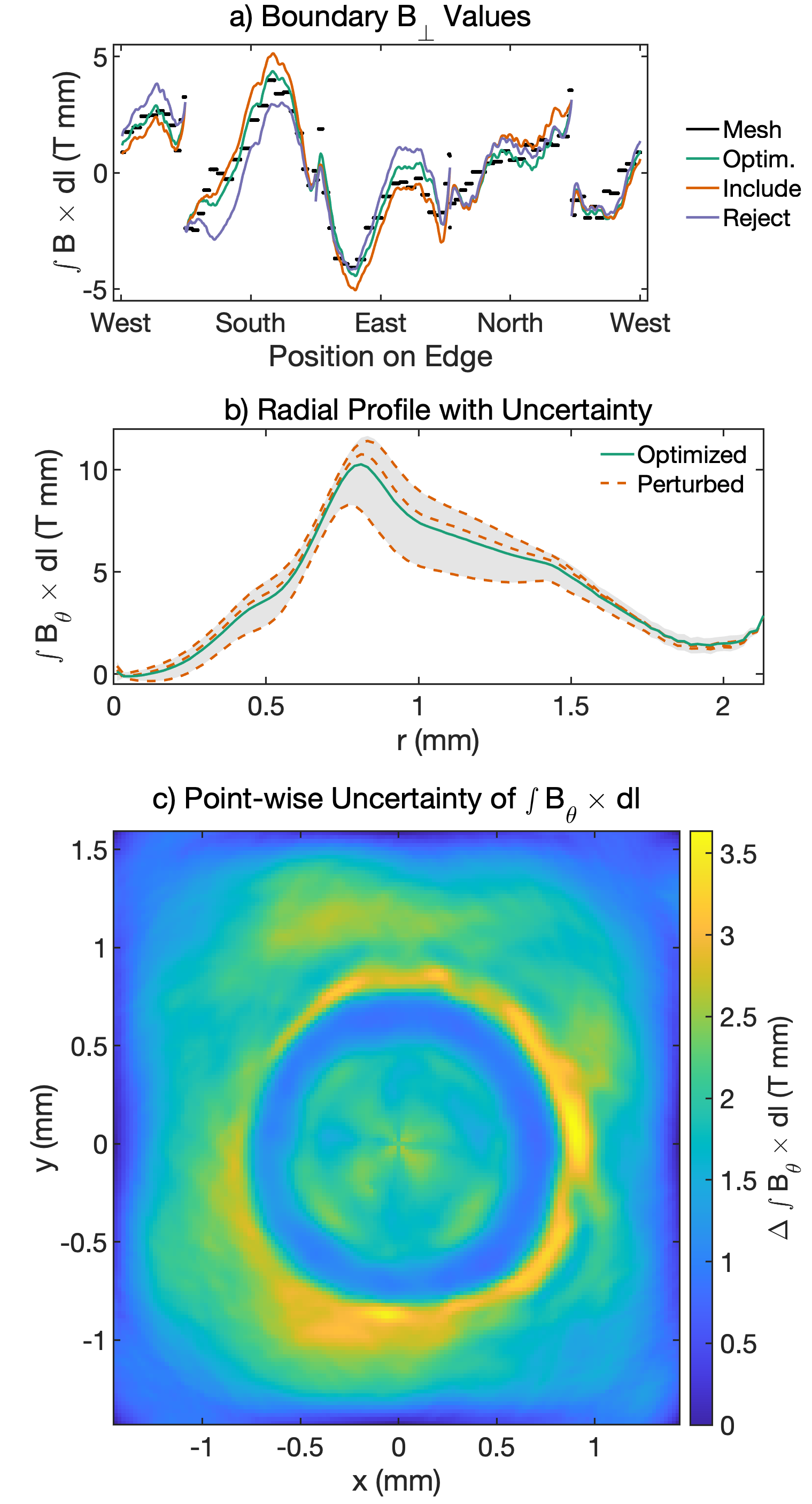}
    \caption{\textbf{Inversion uncertainty for outer mesh}: a) Boundary $B_\perp$ values for the mesh (black), an optimized inversion (green), a weakly perturbed inversion that is included in the ensemble (orange), and a strongly perturbed inversion that is rejected from the ensemble (purple). b) Radial profile of the path-integrated, azimuthal magnetic field with 95\% confidence error bands given by the shaded gray area. The profile of the optimized inversion is shown in green and examples inversion profiles from perturbed sources are shown in dashed orange lines. c) Point-wise uncertainty of the path-integrated, azimuthal magnetic field. Each point is calculated as half the range of values across the ensemble.
}
    \label{fig:EnsembleError102473}
\end{figure}
    

We generate a large ensemble of inversions from randomly generated source profiles, which can be analyzed statistically to estimate the inversion uncertainty. Key distinctions in our approach include 1) using a single optimized source as a baseline for perturbations, and 2) limiting the magnitude of the perturbations to maintain good matching of the boundary information.

To generate each inversion, the coefficient matrix $A$ is perturbed away from its optimal value $\hat{A}$ and the inversion is recalculated. The boundary values of $B_\parallel$ remain matched with the mesh measurements, but the $B_\perp$ error will inevitably increase as quantified by $\chi^2$ in Eqs. \eqref{eq:Chi2} and \eqref{eq:DeltaChi2}. Here, the uncertainty in the mesh measurement $\sigma$ is taken to be 1 T$\,$mm which is approximately one pixel worth of beamlet deflection.

\begin{align}
    \chi^2(A) = \sum_{beamlets} \left[ \frac{(B_{\perp,inv}(A)-B_{\perp,mesh})^2}{\sigma^2} \right] \label{eq:Chi2} \\
    \Delta \chi^2(A) = \chi^2(A) - \chi^2(\hat{A}) \label{eq:DeltaChi2}
\end{align}

If the error in $B_\perp$ grows too large, the perturbed source is unlikely to represent the experimental profile and the inversion is excluded from the ensemble. As we adjust the 15 coefficients in $A$, excluding the $A_{00}$ element, the increase in $\chi^2$ follows the $\chi^2$ probability distribution with 15 degrees of freedom \cite{bevington_data_2003}. Therefore, we can reject any inversions that produce a $\Delta \chi^2$ exceeding the threshold value, which is set based on the desired confidence level and the number of degrees of freedom. A threshold value of $\Delta \chi^2=25$ is used, corresponding to a 95\% confidence interval for 15 degrees of freedom. This approach maps out the source parameter space that encloses 95\% of the probability \cite{bevington_data_2003}. Over 1000 different inversions were calculated and 500 of these had sufficiently low boundary errors to be added to the ensemble.

Boundary values of $B_\perp$ for the mesh and different inversions is shown in Fig. \ref{fig:EnsembleError102473}a. The mesh is displayed by the black curve, the optimized source is in green, and two selected inversions are in orange and purple. The optimized inversion matches the mesh well, while the perturbed inversions deviate. The orange curve has a $\Delta \chi^2$ value of  18 and is added to the ensemble, whereas the purple line accumulates an error of $\Delta \chi^2=$34 and is therefore rejected. Visual inspection of the curves supports this threshold value as the rejected inversion has several regions of correlated errors of 2 T$\,$mm, which is $\sim$50\% of the measured value. The included inversion has boundary values that are only moderately mismatched.



The resulting ensemble can then be used to analyze the inversion uncertainty. For example, each inversion may be azimuthally averaged to obtain a family of radial profiles. The range of these profiles is shown by the gray area in Fig. \ref{fig:EnsembleError102473}b and traces out the 95\% confidence interval from uncertainty in the source optimization. The optimized source is shown in the green curve and several example profiles are shown in dashed orange curves. The half-range has a maximum value of 2.2 T$\ $mm at the peak of the profile located at $r=0.85$ mm, which corresponds to an uncertainty of 21\%. The uncertainty is most pronounced here as different inversions peak at slightly different locations, which leads to a broad range. Furthermore, this region is highly sensitive to source perturbations due to the large field strength and the small spatial scale that the field is peaked over.

Figure \ref{fig:EnsembleError102473}c shows the half-range of the path-integrated azimuthal magnetic field across the ensemble. Again, the uncertainty is peaked around the field maximum at $r=0.85$ mm and has localized pockets of large uncertainty up to 3.5 T$\ $mm. The variation is minimized at the edges where the fields do not change from inversion to inversion significantly due to the enforcement of $B_\parallel$ throughout the ensemble and the small deviations of $B_\perp$ resulting from perturbing the source only slightly. Here, uncertainty in the boundary conditions would contribute and can be included straightforwardly via an additional Monte Carlo analysis. However, we find that the interior of the inversion is relatively insensitive to non-correlated boundary perturbations and has therefore been neglected in this work.

In addition to a point-wise comparison, it is also helpful to directly compare specific features of the different inversions. For example, the magnetic flux in the inversion (defined as $\Phi=\int dz\,dr\,B_\theta$) varies from 6.5 to 12.0 T$\ $mm$^2$ across the ensemble and the location of the peak is bounded from $r=$0.75 to 0.88 mm. As another example, the mechanisms in magnetic reconnection strongly depend on current sheet parameters such as the peak current and sheet width. However, the locations of these features are sensitive to the boundary information and the source profile, which might lead to large point-wise variation across an ensemble. For example, Ref. \cite{fox_proton_2023} shows (in the 1-D case) how variation in the source profile leads to different locations of a magnetic reconnection current sheet in various reconstructions. However, other quantities such as the peak current are relatively robust between reconstructions. For this reason, it will likely be of interest to develop analyses that handle the full family of reconstructions in a sophisticated manner, beyond reducing the data sets to a pointwise mean and variance.


Lastly, using a finite basis representation contributes uncertainty in the form of a truncation error from neglecting higher order terms. A discussion of this effect is given in Appendix \ref{App:TruncErr}.

\section{Conclusion}
In this study, we address the critical issues of correctly determining magnetic boundary conditions and proton source profiles for accurate proton radiography inversions. These factors are often overlooked, primarily due to limitations in measurement capabilities. Here, nonzero boundary conditions were implemented and validated in the PROBLEM code implementation \cite{bott_proton_2017} of the Monge-Amp\`ere inversion algorithm. These boundary conditions were then applied to proton radiography experiments of laser-driven foils, whose fields were reconstructed with both fluence and high-fidelity mesh techniques in the same shot. The mesh measurements were used as a benchmark and provided nonzero boundary conditions for the fluence inversions. We find good agreement between mesh and fluence reconstructions only when nonzero boundary conditions are used.

Furthermore, we introduced a scheme to determine the proton source pattern, a long-standing issue for proton radiography inversions. By embedding a fluence inversion within a mesh region, the mesh provides over-constrained magnetic boundary conditions ($B_\parallel$ and $B_\perp$), which are used to constrain the source. In particular the boundary condition for $B_\parallel$ is applied to coincide with the mesh measurement and the source profile is iteratively tuned to also match the mesh $B_\perp$ measurement. A simple gradient descent approach was used to optimize the source and led to a small error between mesh and fluence boundary values ($B_\perp$ RMSE = 0.7 T$\, $mm). Consequently, the field lines were nearly continuous across the fluence-mesh interface. This scheme significantly improves the accuracy of fluence-based inversions. The repercussions of choosing an incorrect source are also discussed. 

Lastly, we estimated the inversion error stemming from uncertainties in the optimization of the source. We generated an ensemble of inversions using sources that have been perturbed away from the optimized profile. The total inversion uncertainty is calculated from the range of possible inversions that match the boundary information to within a threshold value.

The techniques for boundary condition implementation, source profile optimization, and uncertainty estimation have direct applications for future proton inversions that have boundary information. Future work will directly use these methods to study magnetic reconnection, where boundary conditions and accurate source profiles are especially important to reliably recover the reconnected magnetic flux and reconstruct small-scale structures \cite{fox_fast_2020}. 

\begin{acknowledgments}
The authors thank the OMEGA staff for their help in conducting these experiments and General Atomics for target fabrication. This work was supported by the Department of Energy under grant Nos. DE-NA0004034 and DE-NA0003868 and by the Department of Energy [National Nuclear Security Administration] University of Rochester ``National Inertial Confinement Fusion Program” under Award Number DE-NA0004144. This material is based upon work supported by the National Science Foundation Graduate Research Fellowship Program under Grant No. 2039656. Any opinions, findings, and conclusions or recommendations expressed in this material are those of the author(s) and do not necessarily reflect the views of the National Science Foundation.

This report was prepared as an account of work sponsored by an agency of the United States Government. Neither the United States Government nor any agency thereof, nor any of their employees, makes any warranty, express or implied, or assumes any legal liability or responsibility for the accuracy, completeness, or usefulness of any information, apparatus, product, or process disclosed, or represents that its use would not infringe privately owned rights. Reference herein to any specific commercial product, process, or service by trade name, trademark, manufacturer, or otherwise does not necessarily constitute or imply its endorsement, recommendation, or favoring by the United States Government or any agency thereof. The views and opinions of authors expressed herein do not necessarily state or reflect those of the United States Government or any agency thereof.
\end{acknowledgments}

\appendix 
\section{E vs. B Analysis in Outer Mesh} \label{App:EvB}
Both mesh and fluence methods extract proton deflection, which can originate from electric or magnetic fields. Although the transverse electric field is often subdominant in laser-foil systems, it is important to distinguish between the two fields. One method of decoupling the fields is through the use of two mono-energetic populations of protons, specifically the 3 MeV protons from the D-D fusion reaction and the 14.7 MeV protons from the D-$^3$He fusion reaction as detailed in \cite{griff-mcmahon_measurements_2023}. Measuring the deflection of each proton population breaks the degeneracy of the fields and allows the extraction of path-integrated electric and magnetic fields.

\begin{figure}
    \includegraphics[width=\linewidth]{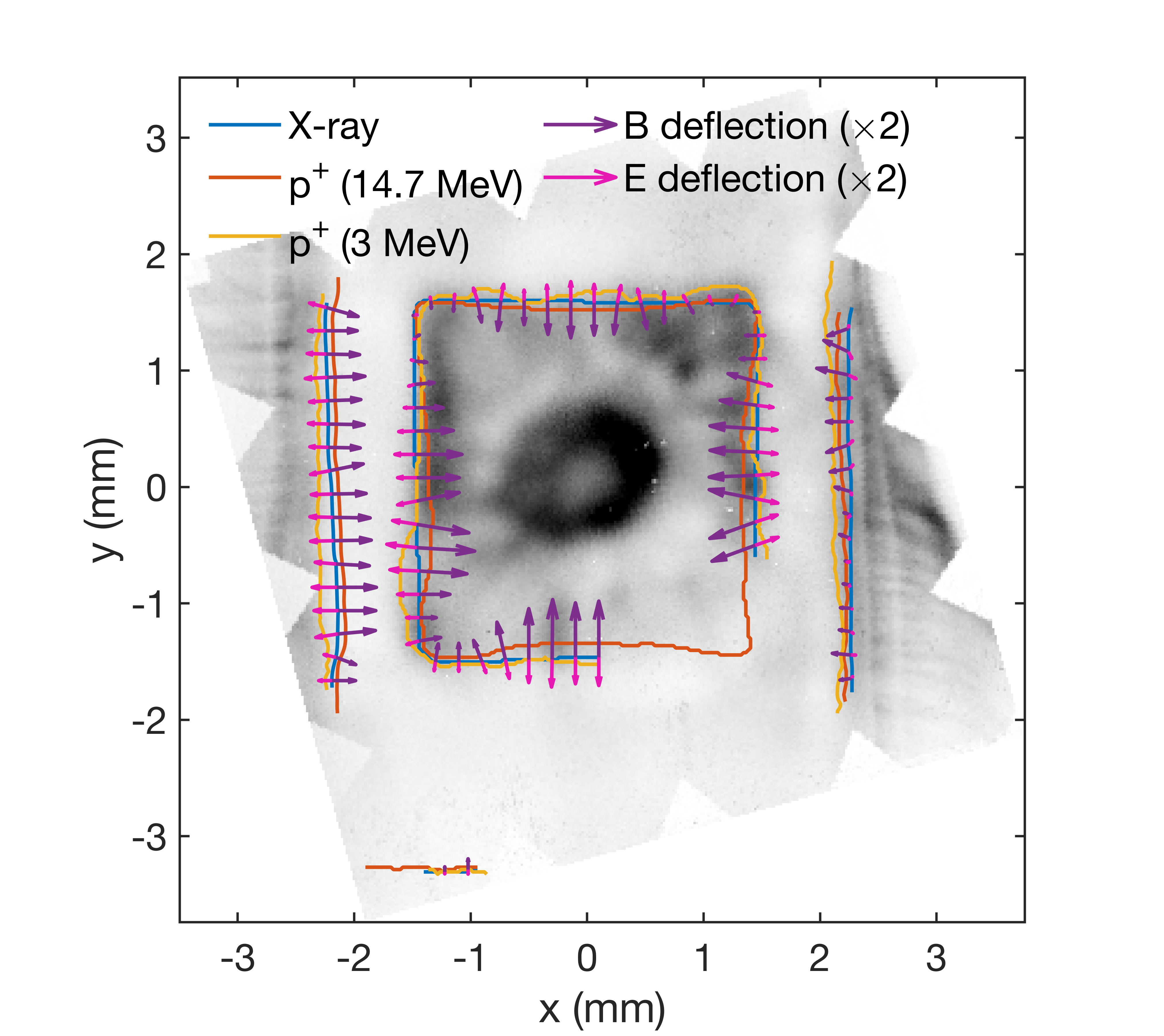}
    \caption{3 MeV proton fluence in outer mesh configuration with overlaid edges of the mesh for x-rays (blue), 14.7 MeV protons (red), and 3 MeV protons (yellow). The magnetic and electric deflections for the 14.7 MeV protons are shown in purple and pink, respectively. Their magnitudes have been scaled 2$\times$ for visibility.}
    \label{fig:EdgeDef73}
\end{figure}

\begin{figure}
    \includegraphics[width=\linewidth]{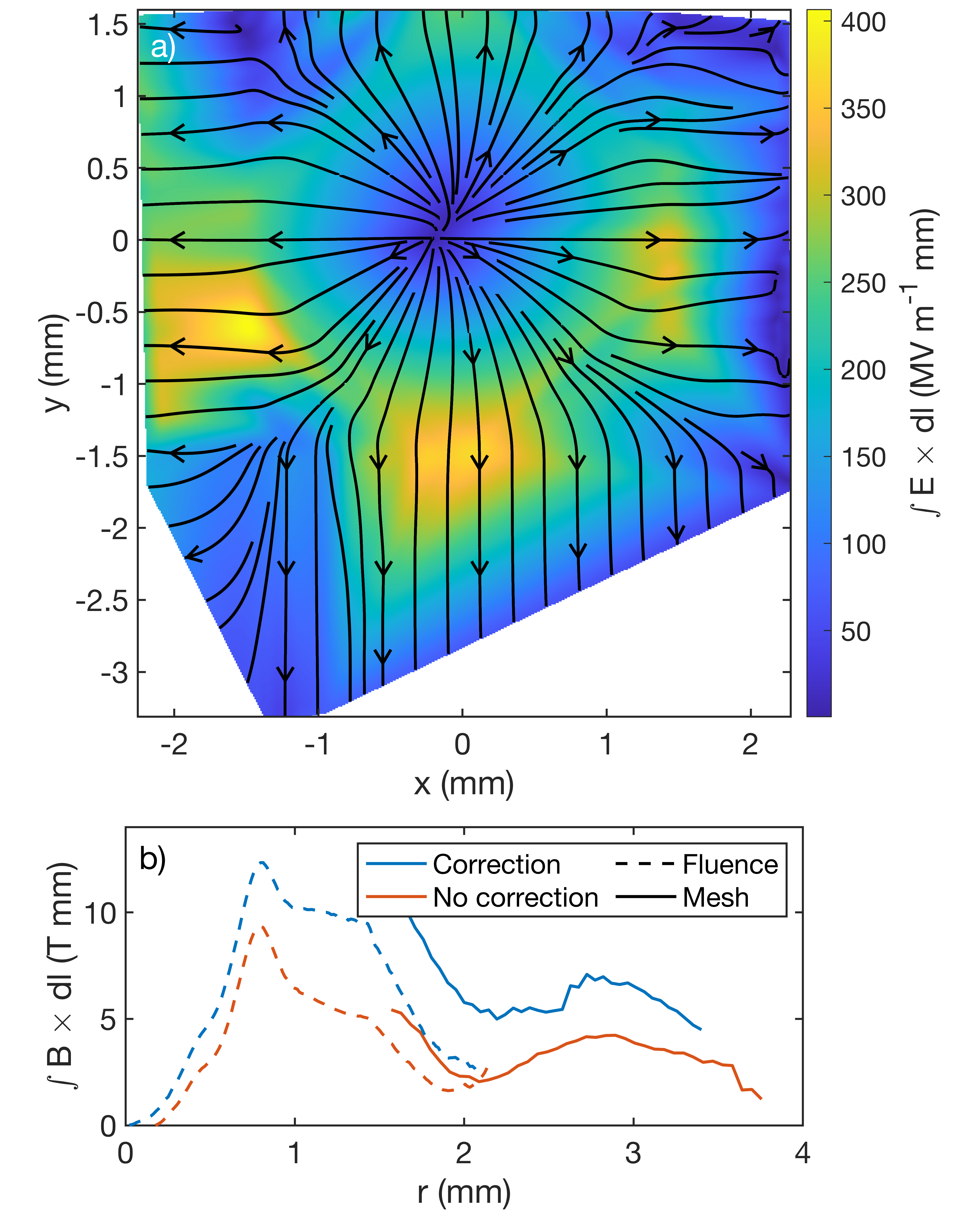}
    \caption{a) Path-integrated electric field after interpolation. b) Radial profile of the path-integrated toroidal magnetic field with electric field correction (blue) and without (red). The dashed lines used fluence inversions and the solid lines used mesh measurements}
    \label{fig:EfieldInterp73}
\end{figure}

Figure \ref{fig:EdgeDef73} shows the D-D proton fluence in the outer mesh configuration. Although the mesh beamlets are blurred due to proton scattering within the target, prominent features like the edges of the mesh are still visible. Therefore, comparison of the edge deflections gives the contribution of the deflection due to electric and magnetic fields. The mesh edges for the x-rays and the two proton populations have been identified and are overlaid in solid lines. Importantly, we find that the edge deflection does not scale with proton energy as $d\propto K^{-1/2}$ as one would expect if the deflection were solely due to magnetic fields. Instead, the deflection scaling indicates the presence of electric fields too.

The magnetic and electric contribution to the D-$^3$He proton deflections is shown by the purple and pink vectors. There is a prominent outward electric field around the inner edge of the mesh that likely explains why the magnitude of the D-$^3$He proton deflections were reduced relative to the other mesh configurations; the electric field partially compensated for the inward magnetic deflection. Note that this method captures only the normal component of the deflection and the 200 ps time-of-flight difference is assumed negligible on the 1 ns field evolution timescale. 

Figure \ref{fig:EfieldInterp73}a shows the path-integrated electric field after linearly interpolating between the edges. Regions outside of the convex hull of the interpolation have been excluded. 
The direction of the electric field is radially outward, which suggests that the electric field near the center of the fluence region is relatively weak. The electric field may originate from charging of the mesh or target stalk, positioned in the upper-right corner of the target. Future efforts may use a mesh with larger pitch to prevent beamlet blurring and would improve the accuracy and coverage of this method.

Compensating for the electric fields significantly alters the magnetic field profile as shown in Fig. \ref{fig:EfieldInterp73}b. The corrected curve more closely matches the other experimental configurations shown in Fig. \ref{fig:Bth_radial}.

\section{Algorithm for Source Optimization Using Gradient Descent} \label{App:GradDescAlg}

\begin{figure} [H]
\begin{algorithm}[H]
    \caption{Source optimization with Grad. Descent}
    \label{alg:sourceOpt}
    \begin{algorithmic} [1]
        \State{Load proton image $P$}
        \State{Define finite difference step size $\lambda\sim0.02$, gradient descent step size $\gamma\sim0.02$, and error threshold $\sim1$ T$\,$mm }
        \If{$P$ is large}
            \State Scale $P$ down to $\sim50\times50$ array
        \EndIf
        \State{Initialize weight array $A_{ij}^{(0)}=\delta_{i0}\delta_{j0}$}
        \State{Generate source $\Psi_0(A^{(0)})$ from Chebyshev expansion of $A^{(0)}$ and normalize to proton fluence in $P$}
        \State{Call inversion until fully converged}
        \State{Calculate RMSE for $B_\perp$ on the boundary: $\epsilon^{(0)}$}
        \For{$i=1$ to $\infty$}
            \ForAll{elements $j>0$ in $A$}
                \State Increase $j^{th}$ element of A and generate $\Psi_0(A^+)$:
                $A^+ = A^{(i-1)}$, $A_j^+ = A^+_j + \lambda$
                \State Call inversion and save RMSE: $\epsilon^+$
                \State Decrease $j^{th}$ element of A and generate $\Psi_0(A^-)$: $A^- = A^{(i-1)}$, $A^-_j = A^-_j - \lambda$
                \State Call inversion and save RMSE: $\epsilon^-$
                \State Calculate derivative: $\partial \epsilon/\partial j \approx (\epsilon^+ - \epsilon^-)/2\lambda$
            \EndFor
            \State Step forward weight array using gradient descent: $A^{(i)}=A^{(i-1)} - \gamma \nabla_j \epsilon$
            \State Generate source $\Psi_0(A^{(i)})$ and call inversion
            \State Calculate and save RMSE: $\epsilon^{(i)}$
            \If{$\epsilon^{(i)}<\theta$ error threshold OR converged}
                \State{BREAK}
            \EndIf
        \EndFor
    \end{algorithmic}
\end{algorithm}
\end{figure}

\section{Null Shot Fluence Variation} \label{App:nullShot}
\begin{figure}
    \includegraphics[width=\linewidth]{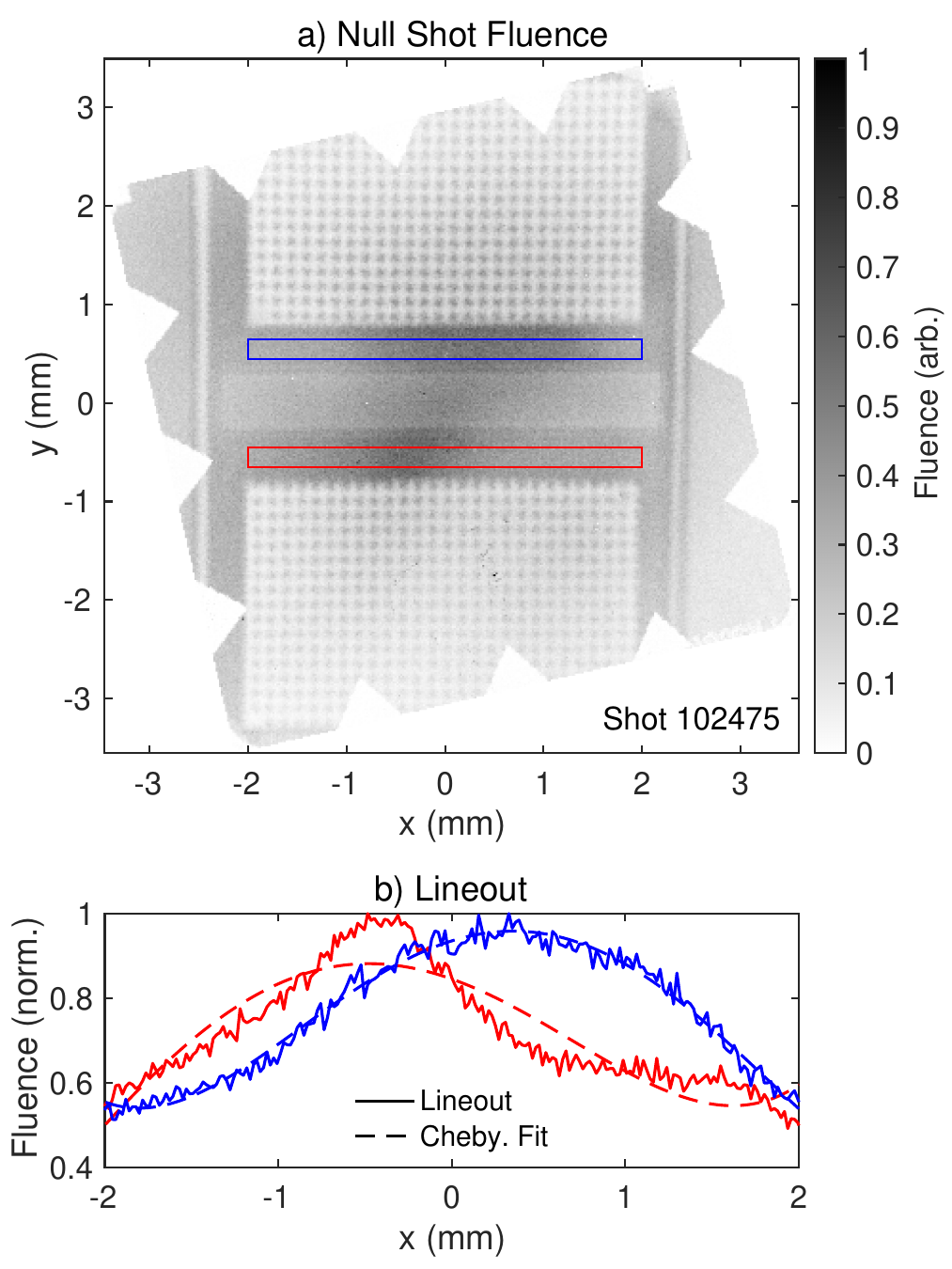}
    \caption{a) 14.7 MeV proton fluence for null shot without drive beams. The blue and red rectangles show where lineouts where taken. b) Fluence lineout averaged in the y-direction over the blue and red regions (solid lines) and a 4th-order Chebyshev polynomial fit (dashed lines).}
    \label{fig:nullShot}
\end{figure}

In order to give confidence in the extracted source profiles, we conducted an undriven ``null" shot to examine the intrinsic fluence variation in the absence of any driven EM fields. Figure \ref{fig:nullShot}a displays the 14.7 MeV proton fluence for a null shot taken under the same backlighter laser configuration as the experiments discussed in Section \ref{sec:Exp}. The target features a mesh on the top and bottom regions as well as a rectangular cutout in the center; these prevent a study of the proton variation over the entire proton image. Instead, we are limited to the regions outlined by the blue and red rectangles which have no target features.

Although shot-to-shot variations preclude a direct comparison between this null shot and the extracted sources, more general statements may be made on the scale and magnitude of the spatial variation. Figure \ref{fig:nullShot}b shows fluence lineouts that have been averaged over the blue and red rectangles. The fluence varies by a factor of 2 over 2 mm in the target plane. The scale and magnitude of these variations is in good agreement with the extracted source profile in the outer mesh configuration discussed in Section \ref{sec:InvSource}. In that configuration, the extracted profile varies by a factor of $\sim$2.5 over 2 mm. Furthermore, a 4th-order Chebyshev polynomial fit (5 terms) is shown in dashed lines and has overall good agreement to the lineouts. This 1-d check supports the use of the Chebyshev basis since agreement is found within a relatively low order. Overall, these agreements lend credibility to our extracted profiles and methodology. 

\section{Source Truncation Error} \label{App:TruncErr}
A sufficiently large basis set is necessary to adequately capture variations in the proton source profile. Neglecting higher-order terms may lead to a poor optimization that induce large boundary errors. Here, we discuss the qualifications for truncating the higher order terms and determine the error in doing so.

\begin{figure}
    \includegraphics[width=\linewidth]{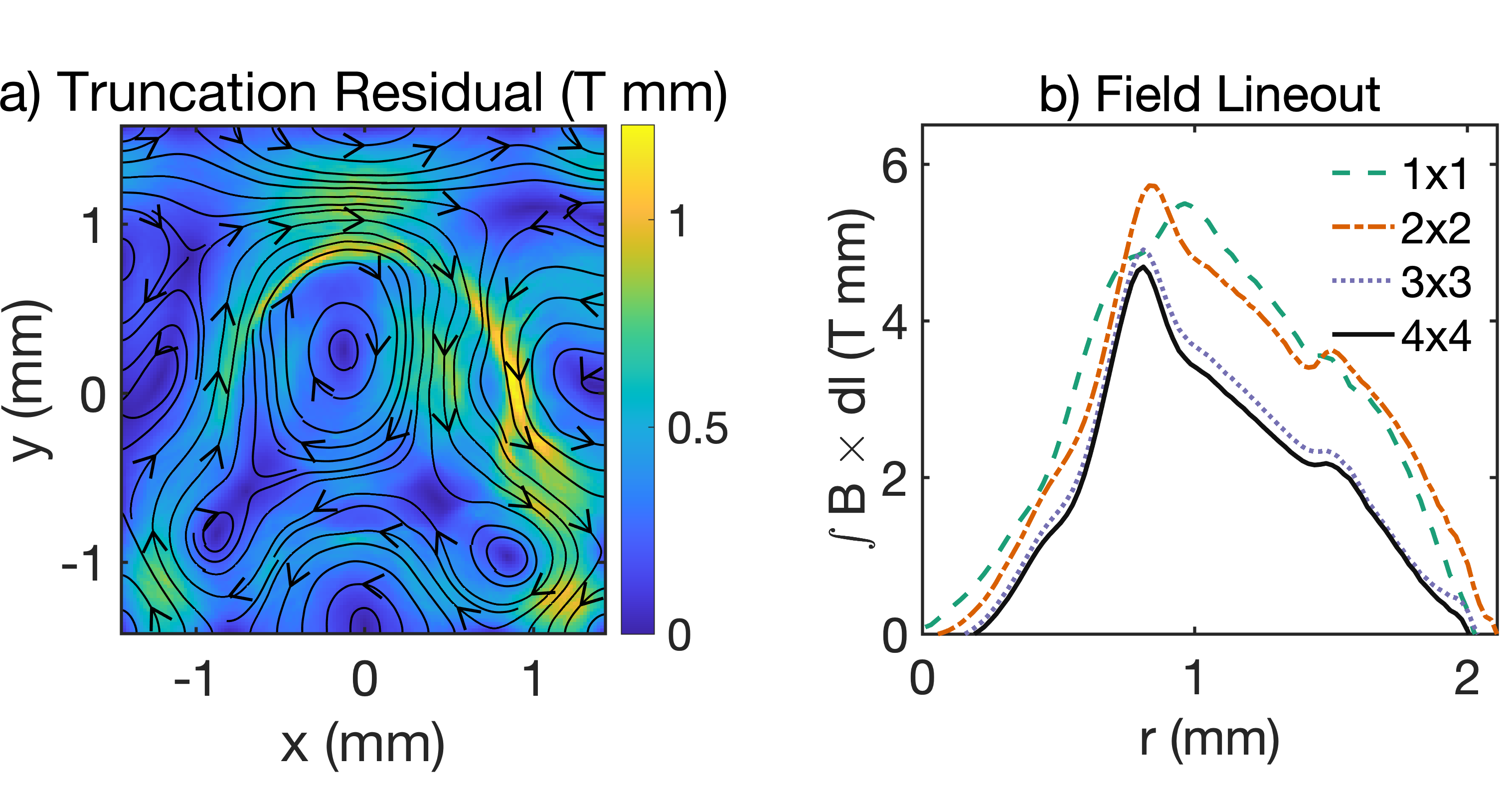}
    \caption{a) Difference in path-integrated magnetic field when a 5$\times$5 versus 4$\times$4 weight array was used to determine the source. b) Toroidal magnetic field profile after azimuthally averaging for different basis matrix sizes. Note that electric field corrections have not been included in this plot.}
    \label{fig:TruncErr}
\end{figure}

The effect of neglecting higher-order terms is studied by optimizing the outer mesh source profile with five basis sets of varying size, ranging from a 1$\times$1 weight array (uniform fluence) to 5$\times$5 (up to 4rd order Chebyshev polynomial in each direction). Figure \ref{fig:TruncErr}a shows the residual of the path-integrated magnetic field associated with truncating the basis at a 4$\times$4 matrix, compared to extending the basis to 5$\times$5. The field profile is largely unchanged between the two basis sets and the inversion residual has a mean value of 0.4 T$\,$mm and maximum value of 1.2 T$\,$mm. These values are small compared to the peak field of $\sim$ 5 T$\,$mm. Note that this is the peak field without correcting for electric fields as outlined in Appendix \ref{App:EvB}. Figure \ref{fig:TruncErr}b shows the radial profile of the different solves and the clear convergence to the final profile, without electric field corrections. The appropriate truncation point can also be estimated using the $B_\perp$ RMSE plateau. The RMSE decreases from 4.8 to 0.7 T$\,$mm as the basis size is increased from 1$\times$1 to 4$\times$4 (used in this paper). Further increasing the size to 5$\times$5 only slightly decreases the RMSE to 0.6 T$\,$mm, indicating that a 4$\times$4 basis is sufficient to minimize the boundary error. The basis set is sufficiently large when both the $B_\perp$ RMSE and the inversions start to converge; low-order terms are necessary to establish the general structure of the source, but higher-order corrections become increasingly unimportant. 



\bibliography{Prad_PROB}

\end{document}